\documentclass[aps,prb,twocolumn,showpacs]{revtex4}

\usepackage{graphicx}
\usepackage{amsmath}
\usepackage{amssymb}

\begin{document}

\title{Two gaps with one energy scale in cuprate superconductors}

\author{Shiping Feng$^{*}$, Huaisong Zhao, and Zheyu Huang}
\affiliation{Department of Physics, Beijing Normal University, Beijing 100875, China}

\begin{abstract}
The interplay between the superconducting gap and normal-state pseudogap in cuprate superconductors is studied based on
the kinetic energy driven superconducting mechanism. It is shown that the interaction between charge carriers and spins directly
from the kinetic energy by exchanging spin excitations induces the normal-state pseudogap state in the particle-hole channel and
superconducting-state in the particle-particle channel, therefore there is a coexistence of the superconducting gap and
normal-state pseudogap in the whole superconducting dome. This normal-state pseudogap is closely related to the quasiparticle
coherent weight, and is a necessary ingredient for superconductivity in cuprate superconductors. In particular, both the
normal-state pseudogap and superconducting gap are dominated by one energy scale, and they are the result of the strong electron
correlation.
\end{abstract}


\maketitle

After intensive investigations over more than two decades, it has become clear that the interplay between the superconducting
(SC) gap and normal-state pseudogap is one of the most important problems in cuprate superconductors
\cite{Hufner08,Timusk99}. The parent compounds of cuprate superconductors are believed to belong to a class of materials known
as Mott insulators with an antiferromagnetic (AF) long-range order (AFLRO), where a single common feature is the presence of the
CuO$_{2}$ plane \cite{Damascelli03}. As the CuO$_{2}$ planes are doped with charge carriers, the AF phase subsides and
superconductivity emerges \cite{Damascelli03}, then the physical properties mainly depend on the extent of doping, and the
regimes have been classified into the underdoped, optimally doped, and overdoped, respectively. Experimentally, a large body
of experimental data available from a wide variety of measurement techniques have provided rather detailed information on the
normal-state pseudogap state and SC-state in cuprate superconductors
\cite{Hufner08,Timusk99,Damascelli03,Deutscher05,Eschrig06,Devereaux07,Fischer07,Batlogg94,Loeser96,McElroy05,Tacon06,Dai01},
where an agreement has emerged that the normal-state pseudogap state is particularly obvious in the underdoped regime, i.e.,
the magnitude of the normal-state pseudogap is much larger than that of the SC gap in the underdoped regime, then it smoothly
decreases upon increasing the doping concentration. This is why the normal-state properties in the underdoped regime exhibit a
number of the anomalous properties \cite{Kastner98}. However, there is a controversy about the phase diagram with respect to the
normal-state pseudogap line \cite{Hufner08}. On the one hand, some authors \cite{Cho06} analyzed the experimental data, and then
they argued that the normal-state pseudogap line intersects the SC dome at about optimal doping. On the other hand, it has been
argued that the normal-state pseudogap merges gradually with the SC gap in the overdoped regime \cite{Millis06}, eventually
disappearing together with superconductivity at the end of the SC dome. Theoretically, to the best of our knowledge, all theoretical
studies of the normal-state pseudogap phenomenon and its relevance to superconductivity performed so far are based on the
phenomenological d-wave Bardeen-Cooper-Schrieffer (BCS) formalism \cite{Cho06,Millis06,Hufner08b,Norman05,Benfatto00}, where
by introducing a phenomenological doping and temperature dependence of the normal-state pseudogap, the two-gap feature in cuprate
superconductors is reproduced \cite{Benfatto00}. In particular, a phenomenological theory of the normal-state pseudogap state has
been developed \cite{Yang06}, where an ansatz is proposed for the coherent part of the single particle Green's function in a doped
resonant valence bond state, and then the calculated result of the electronic properties in the normal-state pseudogap phase is in
qualitative agreement with the experimental data. Moreover, it has been argued recently that the pseudogap is a combination of a
quantum disordered d-wave superconductor and an entirely different form of competing order, originating from the particle-hole
channel \cite{Tesanovic08}. However, up to now, the interplay between the SC gap and normal-state pseudogap in cuprate
superconductors has not been treated starting from a microscopic SC theory, therefore no general consensus for the normal-state
pseudogap has been reached yet on the its origin, its role in the onset of superconductivity itself, and not even on its evolution
across the phase diagram of cuprate superconductors \cite{Hufner08}.

In our earlier work, a kinetic energy driven SC mechanism has been developed \cite{feng0306}, where the interaction between
charge carriers and spins directly from the kinetic energy by exchanging spin excitations induces a d-wave charge carrier pairing
state, and then their condensation reveals the SC ground-state. In particular, this SC-state is controlled by both the SC gap and
quasiparticle coherence, which leads to that the maximal SC transition temperature occurs around the optimal doping, and then
decreases in both the underdoped and overdoped regimes. Within this kinetic energy driven SC mechanism, we have discussed the low
energy electronic structure \cite{guo07}, quasiparticle transport \cite{wang08}, and Meissner effect \cite{feng10}, and qualitatively
reproduced some main features of the corresponding experimental results of cuprate superconductors in the SC-state. In this paper, we
study the interplay between the SC gap and normal-state pseudogap in cuprate superconductors based on this kinetic energy driven SC
mechanism \cite{feng0306}, where one of our main results is that the interaction between charge carriers and spins directly from the
kinetic energy by exchanging spin excitations induces the normal-state pseudogap state in the particle-hole channel and SC-state in
the particle-particle channel, therefore there is a coexistence of the SC gap and normal-state pseudogap in the whole SC dome. Our
results also show that both the normal-state pseudogap and SC gap are dominated by one energy scale, and the normal-state pseudogap
also is a necessary ingredient for superconductivity.

In cuprate superconductors, the characteristic feature is the presence of the CuO$_{2}$ plane \cite{Damascelli03}. In this case,
it is commonly accepted that the essential physics of the doped CuO$_{2}$ plane \cite{anderson87} is captured by the $t$-$J$ model
on a square lattice,
\begin{eqnarray}\label{tjham}
H&=&-t\sum_{i\hat{\eta}\sigma}C^{\dagger}_{i\sigma}C_{i+\hat{\eta}\sigma}+t'\sum_{i\hat{\tau}\sigma}C^{\dagger}_{i\sigma}
C_{i+\hat{\tau}\sigma}\nonumber\\
&+&\mu\sum_{i\sigma} C^{\dagger}_{i\sigma}C_{i\sigma}+J\sum_{i\hat{\eta}}{\bf S}_{i}\cdot
{\bf S}_{i+\hat{\eta}},
\end{eqnarray}
where $\hat{\eta}=\pm\hat{x},\pm\hat{y}$, $\hat{\tau}=\pm\hat{x}\pm\hat{y}$, $C^{\dagger}_{i\sigma}$ ($C_{i\sigma}$) is the
electron creation (annihilation) operator, ${\bf S}_{i}=(S^{x}_{i},S^{y}_{i},S^{z}_{i})$ are spin operators, and $\mu$ is the
chemical potential. This $t$-$J$ model (\ref{tjham}) is subject to an important local constraint
$\sum_{\sigma}C^{\dagger}_{i\sigma}C_{i\sigma} \leq 1$ to avoid the double occupancy. To incorporate this electron single
occupancy local constraint, the charge-spin separation (CSS) fermion-spin theory \cite{feng04} has been proposed, where the
physics of no double occupancy is taken into account by representing the electron as a composite object created by
$C_{i\uparrow}=h^{\dagger}_{i\uparrow}S^{-}_{i}$ and $C_{i\downarrow}=h^{\dagger}_{i\downarrow}S^{+}_{i}$, with the spinful
fermion operator $h_{i\sigma}=e^{-i\Phi_{i\sigma}}h_{i}$ that describes the charge degree of freedom of the electron together with
some effects of spin configuration rearrangements due to the presence of the doped hole itself (charge carrier), while the spin
operator $S_{i}$ represents the spin degree of freedom of the electron, then the electron single occupancy local constraint is
satisfied in analytical calculations. In this CSS fermion-spin representation, the $t$-$J$ model (\ref{tjham}) can be expressed as
\cite{feng0306,feng04},
\begin{eqnarray}\label{cssham}
H&=&t\sum_{i\hat{\eta}}(h^{\dagger}_{i+\hat{\eta}\uparrow}h_{i\uparrow}S^{+}_{i}S^{-}_{i+\hat{\eta}}+
h^{\dagger}_{i+\hat{\eta}\downarrow}h_{i\downarrow}S^{-}_{i}S^{+}_{i+\hat{\eta}})\nonumber\\
&-&t'\sum_{i\hat{\tau}}(h^{\dagger}_{i+\hat{\tau}\uparrow}h_{i\uparrow}S^{+}_{i}S^{-}_{i+\hat{\tau}}+
h^{\dagger}_{i+\hat{\tau}\downarrow}h_{i\downarrow}S^{-}_{i}S^{+}_{i+\hat{\tau}})\nonumber\\
&-&\mu\sum_{i\sigma} h^{\dagger}_{i\sigma}h_{i\sigma}+J_{{\rm eff}}\sum_{i\hat{\eta}}{\bf S}_{i}\cdot {\bf S}_{i+\hat{\eta}},
\end{eqnarray}
where
$J_{{\rm eff}}=(1-\delta)^{2}J$, and $\delta=\langle h^{\dagger}_{i\sigma}h_{i\sigma}\rangle=\langle h^{\dagger}_{i}h_{i}\rangle$
is the doping concentration. As a consequence, the kinetic energy in the $t$-$J$ model has been transferred as the interaction
between charge carriers and spins, which reflects that even the kinetic energy in the $t$-$J$ model has strong Coulombic
contribution due to the restriction of single occupancy of a given site.

As in the conventional superconductors, the SC-state in cuprate superconductors is also characterized by the electron Cooper pairs,
forming SC quasiparticles \cite{tsuei00}. On the other hand, the angle resolved photoemission spectroscopy measurements
\cite{shen93,Damascelli03} have shown that in the real space the gap function and pairing force have a range of one lattice spacing.
In this case, the order parameter for the electron Cooper pair can be expressed as \cite{feng0306},
\begin{eqnarray}\label{cooperpair}
\Delta &=&\langle C^{\dagger}_{i\uparrow}C^{\dagger}_{i+\hat{\eta}\downarrow}-C^{\dagger}_{i\downarrow}
C^{\dagger}_{i+\hat{\eta}\uparrow}\rangle \nonumber\\
&=&\langle h_{i\uparrow}h_{i+\hat{\eta}\downarrow}S^{+}_{i} S^{-}_{i+\hat{\eta}}-
h_{i\downarrow}h_{i+\hat{\eta}\uparrow} S^{-}_{i}S^{+}_{i+\hat{\eta}}\rangle.
\end{eqnarray}
In the doped regime without AFLRO, the charge carriers move in the background of the disordered spin liquid state, where the spin
correlation functions $\langle S^{+}_{i}S^{-}_{i+\hat{\eta}}\rangle=\langle S^{-}_{i}S^{+}_{i+\hat{\eta}}\rangle=\chi_{1}$, and
then the SC gap parameter in Eq. (\ref{cooperpair}) can be written as $\Delta=-\chi_{1}\Delta_{\rm h}$, with the charge carrier pair
gap parameter,
\begin{eqnarray}
\Delta_{\rm h}=\langle h_{i+\hat{\eta}\downarrow}h_{i\uparrow}-h_{i+\hat{\eta}\uparrow} h_{i\downarrow}\rangle,
\end{eqnarray}
which shows that the SC gap parameter is closely related to the charge carrier pair gap parameter, therefore the essential physics
in the SC-state is dominated by the corresponding one in the charge carrier pairing state. However, in the extreme low doped regime
with AFLRO, where the spin correlation functions
$\langle S^{+}_{i}S^{-}_{i+\hat{\eta}}\rangle\neq\langle S^{-}_{i}S^{+}_{i+\hat{\eta}}\rangle$, and the conduct is disrupted by
AFLRO. Therefore in the following discussions, we only focus on the case without AFLRO as our previous studies \cite{feng0306}.

The interaction between charge carriers and spins in the $t$-$J$ model (\ref{cssham}) is quite strong, and we \cite{feng0306} have
shown in terms of Eliashberg's strong coupling theory \cite{Eliashberg66} that in the case without AFLRO, this interaction can
induce the d-wave electron Cooper pairing state by exchanging spin excitations. Following our previous discussions \cite{feng0306},
the self-consistent equations that satisfied by the full charge carrier diagonal and off-diagonal Green's functions are obtained as,
\begin{subequations}\label{Green's-function}
\begin{eqnarray}
g(k)&=&g^{(0)}(k)+g^{(0)}(k)[\Sigma^{({\rm h})}_{1}(k)g(k)\nonumber\\
&-&\Sigma^{({\rm h})}_{2}(-k)\Gamma^{\dagger}(k)], \\
\Gamma^{\dagger}(k)&=&g^{(0)}(-k)[\Sigma^{({\rm h})}_{1}(-k)\Gamma^{\dagger}(-k)\nonumber\\
&+&\Sigma^{({\rm h})}_{2}(-k)g(k)],
\end{eqnarray}
\end{subequations}
respectively, where the four-vector notation $k=({\bf k},i\omega_{n})$, and the charge carrier mean-field (MF) Green's function
\cite{feng0306,guo07}, $g^{(0)-1}(k)=i\omega_{n}-\xi_{\bf k}$, with the MF charge carrier spectrum
$\xi_{\bf k}=Zt\chi_{1}\gamma_{{\bf k}}-Zt'\chi_{2}\gamma_{{\bf k}}'-\mu$, the spin correlation function
$\chi_{2}=\langle S_{i}^{+}S_{i+\hat{\tau}}^{-}\rangle$, $\gamma_{{\bf k}}=(1/Z)\sum_{\hat{\eta}}e^{i{\bf k}\cdot\hat{\eta}}$,
$\gamma_{{\bf k}}'=(1/Z)\sum_{\hat{\tau}}e^{i{\bf k}\cdot\hat{\tau}}$, and $Z$ is the number of the nearest neighbor or
next-nearest neighbor sites, while the self-energies $\Sigma^{({\rm h})}_{1}(k)$ in the particle-hole channel and
$\Sigma^{({\rm h})}_{2}(k)$ in the particle-particle channel have been evaluated from the spin bubble as \cite{feng0306},
\begin{subequations}\label{self-energy}
\begin{eqnarray}
\Sigma^{({\rm }h)}_{1}(k)&=&{1\over N^{2}}\sum_{{\bf p,p'}}\Lambda^{2}_{{\bf p}+{\bf p}'+{\bf k}}\nonumber\\
&\times& {1\over \beta}\sum_{ip_{m}}g(p+k)\Pi({\bf p},{\bf p}',ip_{m}), \\
\Sigma^{({\rm h})}_{2}(k)&=&{1\over N^{2}}\sum_{{\bf p,p'}}\Lambda^{2}_{{\bf p}+{\bf p}'+{\bf k}}\nonumber\\
&\times& {1\over \beta}\sum_{ip_{m}}\Gamma^{\dagger}(-p-k)\Pi({\bf p},{\bf p}',ip_{m}),
\end{eqnarray}
\end{subequations}
with $\Lambda_{{\bf k}}=Zt\gamma_{\bf k}-Zt'\gamma_{\bf k}'$, and the spin bubble,
\begin{eqnarray}\label{spin-bubble}
\Pi({\bf p},{\bf p}',ip_{m})={1\over\beta}\sum_{ip'_{m}}D^{(0)}(p')D^{(0)}(p'+p),
\end{eqnarray}
where $p=({\bf p},ip_{m})$, $p'=({\bf p'},ip_{m}')$, and the MF spin Green's function,
$D^{(0)-1}(p)=[(ip_{m})^{2}-\omega_{\bf p}^{2}]/B_{\bf p}$, with the MF spin excitation spectrum $\omega_{\bf p}$ and $B_{\bf p}$
have been given in Ref. \onlinecite{guo07}.

Since the pairing force and charge carrier pair gap have been incorporated into the self-energy $\Sigma^{({\rm h})}_{2}(k)$, it is
called as the effective charge carrier pair gap $\bar{\Delta}_{\rm h}(k)=\Sigma^{({\rm h})}_{2}(k)$. On the other hand, the
self-energy $\Sigma^{({\rm h})}_{1}(k)$ renormalizes the MF charge carrier spectrum \cite{guo06}. In particular,
$\Sigma^{({\rm h})}_{2}(k)$ is an even function of $i\omega_{n}$, while $\Sigma^{({\rm h})}_{1}(k)$ is not. In our previous
discussions \cite{feng0306}, $\Sigma^{({\rm h})}_{1}(k)$ has been broken up into its symmetric and antisymmetric parts as,
$\Sigma^{({\rm h})}_{1}(k)=\Sigma^{({\rm h})}_{\rm 1e}(k)+i\omega_{n}\Sigma^{({\rm h})}_{\rm 1o}(k)$, then both
$\Sigma^{({\rm h})}_{\rm 1e}(k)$ and $\Sigma^{({\rm h})}_{\rm 1o}(k)$ are an even function of $i\omega_{n}$. In this case, we
\cite{feng0306} have defined the charge carrier quasiparticle coherent weight
$Z^{-1}_{\rm hF}(k)=1-{\rm Re}\Sigma^{({\rm h})}_{\rm 1o}(k)$. In the static limit approximation for the effective charge carrier
pair gap and quasiparticle coherent weight, i.e., $\bar{\Delta}_{\rm h}({\bf k})=\bar{\Delta}_{\rm h}\gamma^{({\rm d})}_{{\bf k}}$
with $\gamma^{({\rm d})}_{{\bf k}}=({\rm cos} k_{x}-{\rm cos}k_{y})/2$, and
$Z^{-1}_{\rm hF}=1-{\rm Re}\Sigma^{({\rm h})}_{\rm 1o}({\bf k},\omega=0)|_{{\bf k}=[\pi,0]}$, the BCS-like charge carrier diagonal
and off-diagonal Green's functions with the d-wave symmetry have been obtained \cite{feng0306,guo07}, although the pairing
mechanism is driven by the kinetic energy by exchanging spin excitations. With the help of these charge carrier diagonal and
off-diagonal Green's functions, $\Sigma^{({\rm h})}_{1}(k)$ and $\bar{\Delta}_{\rm h}({\bf k})$ in Eq. (\ref{self-energy}) have
been obtained explicitly as \cite{feng0306,guo07},
\begin{subequations}\label{self-energy-1}
\begin{eqnarray}
\Sigma^{({\rm }h)}_{1}(k)&=&{1\over N^{2}}\sum_{{\bf p}{\bf p}'\mu\nu\tau}\Lambda^{2}_{{\bf p}+{\bf p}'+{\bf k}}
{B_{{\bf p}'}B_{{\bf p}'+{\bf p}}\over 8\omega_{\mu{\bf p}'}\omega_{\nu{\bf p}'+{\bf p}}}\nonumber\\
&\times& {A_{\tau}({\bf p}+{\bf k})F_{\mu\nu\tau}({\bf p},{\bf p}',{\bf k})\over i\omega_{n}+\omega_{\nu{\bf p}'+{\bf p}}-
\omega_{\mu{\bf p}'}-E^{(\tau)}_{{\rm h}{\bf p}+{\bf k}}},~~~~~~~~~\\
\bar{\Delta}_{\rm h}({\bf k})&=&{1\over N^{2}}\sum_{{\bf p}{\bf p}'\mu\nu\tau}
\Lambda^{2}_{{\bf p}+{\bf p}'+{\bf k}}{B_{{\bf p}'}B_{{\bf p}'+{\bf p}}\over 8\omega_{\mu{\bf p}'}
\omega_{\nu{\bf p}'+{\bf p}}}\nonumber\\
&\times& {\bar{\Delta}_{\rm hZ}({\bf p}+{\bf k})\over E^{(\tau)}_{{\rm h}{\bf p}+{\bf k}}}
{F_{\mu\nu\tau}({\bf p},{\bf p}',{\bf k})\over \omega_{\nu{\bf p}'+{\bf p}}-\omega_{\mu{\bf p}'}-
E^{(\tau)}_{{\rm h}{\bf p}+{\bf k}}},~~~~~~~
\end{eqnarray}
\end{subequations}
where $F_{\mu\nu\tau}({\bf p},{\bf p}',{\bf k})=Z_{\rm hF}\{n_{\rm F}(E^{(\tau)}_{{\rm h}{\bf p}+{\bf k}})[n_{\rm B}
(\omega_{\mu{\bf p}'})-n_{\rm B}(\omega_{\nu{\bf p}'+{\bf p}})]+n_{\rm B}(\omega_{\nu{\bf p}'+{\bf p}})[1+n_{\rm B}
(\omega_{\mu{\bf p}'})]\}$, $\mu,\nu,\tau=1,2$, $A_{\tau}({\bf p}+{\bf k})=1+\bar{\xi}_{{\bf p}+{\bf k}}
/E^{(\tau)}_{{\rm h}{\bf p}+{\bf k}}$, $\omega_{1{\bf p}}=\omega_{\bf p}$, $\omega_{2{\bf p}}=-\omega_{\bf p}$,
$E^{(1)}_{{\rm h}{\bf k}}=E_{{\rm h}{\bf k}}$, $E^{(2)}_{{\rm h}{\bf k}}=-E_{{\rm h}{\bf k}}$, the renormalized charge carrier
excitation spectrum $\bar{\xi_{{\bf k}}}=Z_{\rm hF}\xi_{{\bf k}}$, the renormalized charge carrier pair gap
$\bar{\Delta}_{\rm hZ}({\bf k})=Z_{\rm hF}\bar{\Delta}_{\rm h}({\bf k})$, the charge carrier quasiparticle spectrum
$E_{{\rm h}{\bf k}}=\sqrt{\bar{\xi^{2}_{{\bf k}}}+\mid\bar{\Delta}_{\rm hZ}({\bf k})\mid^{2}}$, and
$n_{\rm B}(\omega_{{\bf p}})$ and $n_{\rm F}(E_{h{\bf k}})$ are the boson and fermion distribution functions,
respectively. In this case, the effective charge carrier pair gap parameter can be obtained in terms of Eq. (\ref{self-energy-1})
as \cite{feng0306,guo07},
\begin{eqnarray}\label{hgap}
\bar{\Delta}_{\rm h}={4\over N}\sum_{\bf k}\gamma^{({\rm d})}_{{\bf k}}\bar{\Delta}_{\rm h}({\bf k}).
\end{eqnarray}

However, the self-energy $\Sigma^{({\rm h})}_{1}(k)$ in Eq. (\ref{self-energy-1}) also can be rewritten as,
\begin{eqnarray}\label{self-energy-2}
\Sigma^{({\rm }h)}_{1}(k)&=&{[2\bar{\Delta}_{\rm pg}({\bf k})]^{2}\over i\omega_{n}+M_{\bf k}},
\end{eqnarray}
where $M_{\bf k}$ is the energy spectrum of $\Sigma^{({\rm h})}_{1}(k)$. As in the case of the effective charge carrier pair gap,
since the interaction force and normal-state pseudogap have been incorporated into $\bar{\Delta}_{\rm pg}({\bf k})$, it is called
as the effective normal-state pseudogap. In this case, it is easy to find that in our previous static limit approximation
\cite{feng0306,guo07} for $\Sigma^{({\rm h})}_{\rm 1o}(k)$, the quasiparticle coherent weight
$Z^{-1}_{\rm hF}=\{1+[2\bar{\Delta}_{\rm pg}({\bf k})]^{2}/M^{2}_{\bf k}\}|_{{\bf k}=[\pi,0]}$, which reflects that the partial
effect of the normal-state pseudogap has been contained in the quasiparticle coherent weight. Since the SC-state in the kinetic
energy driven SC mechanism is controlled by both the SC gap and quasiparticle coherence \cite{feng0306}, then in this sense, the
normal-state pseudogap is a necessary ingredient for superconductivity. In the following discussions, we focus on the connection
between the normal-state pseudogap and SC gap beyond our previous static limit approximation \cite{feng0306} for the self-energy
$\Sigma^{({\rm h})}_{1}(k)$, and show explicitly the two-gap feature in cuprate superconductors. Substituting the self-energy
$\Sigma^{({\rm h})}_{1}(k)$ in Eq. (\ref{self-energy-2}) into Eq. (\ref{Green's-function}), we obtain the full charge carrier
diagonal and off-diagonal Green's functions as,
\begin{subequations}\label{Green's-function-1}
\begin{eqnarray}
g(k)&=&{1\over i\omega_{n}-\xi_{\bf k}-\Sigma^{({\rm h})}_{1}(k)-\bar{\Delta}^{2}_{\rm h}({\bf k})/[i\omega_{n}+\xi_{\bf k}
+\Sigma^{({\rm h})}_{1}(-k)]}\nonumber\\
&=&{U^{2}_{1{\rm h}{\bf k}}\over i\omega_{n}-E_{1{\rm h}{\bf k}}}+{V^{2}_{1{\rm h}{\bf k}}\over i\omega_{n}+E_{1{\rm h}{\bf k}}}
\nonumber\\
&+&{U^{2}_{2{\rm h}{\bf k}}\over i\omega_{n}-E_{2{\rm h}{\bf k}}}+{V^{2}_{2{\rm h}{\bf k}}\over i\omega_{n}+E_{2{\rm h}{\bf k}}}, 
\\
\Gamma^{\dagger}(k)&=&{-\bar{\Delta}_{\rm h}({\bf k})\over [i\omega_{n}-\xi_{\bf k}-\Sigma^{({\rm h})}_{1}(k)][i\omega_{n}
+\xi_{\bf k}+\Sigma^{({\rm h})}_{1}(-k)]-\bar{\Delta}^{2}_{\rm h}({\bf k})}\nonumber\\
&=&-{\alpha_{1{\bf k}}\bar{\Delta}_{\rm h}({\bf k})\over 2 E_{1{\rm h}{\bf k}}}\left ({1\over i\omega_{n}-E_{1{\rm h}{\bf k}}}-
{1\over i\omega_{n}+E_{1{\rm h}{\bf k}}}\right )\nonumber\\
&+&{\alpha_{2{\bf k}}\bar{\Delta}_{\rm h}({\bf k})\over 2 E_{2{\rm h}{\bf k}}}\left ({1\over i\omega_{n}-E_{2{\rm h}{\bf k}}}-
{1\over i\omega_{n}+E_{2{\rm h}{\bf k}}}\right ),
\end{eqnarray}
\end{subequations}
where $\alpha_{1{\bf k}}=(E^{2}_{1{\rm h}{\bf k}}-M^{2}_{\bf k})/(E^{2}_{1{\rm h}{\bf k}}-E^{2}_{2{\rm h}{\bf k}})$,
$\alpha_{2{\bf k}}=(E^{2}_{2{\rm h}{\bf k}}-M^{2}_{\bf k})/(E^{2}_{1{\rm h}{\bf k}}-E^{2}_{2{\rm h}{\bf k}})$, and there are four
branches of the charge carrier quasiparticle spectrum due to the presence of the normal-state pseudogap and SC gap,
$E_{1{\rm h}{\bf k}}$, $-E_{1{\rm h}{\bf k}}$, $E_{2{\rm h}{\bf k}}$, and $-E_{2{\rm h}{\bf k}}$, with
$E_{1{\rm h}{\bf k}}=\sqrt{[\Omega_{\bf k}+\Theta_{\bf k}]/2}$, $E_{2{\rm h}{\bf k}}=\sqrt{[\Omega_{\bf k}-\Theta_{\bf k}]/2}$,
and the kernel functions,
\begin{subequations}
\begin{eqnarray}
\Omega_{\bf k}&=&\xi^{2}_{\bf k}+M^{2}_{\bf k}+8\bar{\Delta}^{2}_{\rm pg}({\bf k})+\bar{\Delta}^{2}_{\rm h}({\bf k}),\\
\Theta_{\bf k}&=&\sqrt{(\xi^{2}_{\bf k}-M^{2}_{\bf k})\beta_{1{\bf k}}+16\bar{\Delta}^{2}_{\rm pg}({\bf k})\beta_{2{\bf k}}+
\bar{\Delta}^{4}_{\rm h}({\bf k})},~~~~~~
\end{eqnarray}
\end{subequations}
with $\beta_{1{\bf k}}=\xi^{2}_{\bf k}-M^{2}_{\bf k}+2\bar{\Delta}^{2}_{\rm h}({\bf k})$,
$\beta_{2{\bf k}}=(\xi_{\bf k}-M_{\bf k})^{2}+\bar{\Delta}^{2}_{\rm h}({\bf k})$, while the coherence factors,
\begin{subequations}\label{coherence-factors}
\begin{eqnarray}
U^{2}_{1{\rm h}{\bf k}}&=&{1\over 2}\left [\alpha_{1{\bf k}}\left (1+{\xi_{\bf k}\over E_{1{\rm h}{\bf k}}}\right )-
\alpha_{3{\bf k}}\left (1+{M_{\bf k}\over E_{1{\rm h}{\bf k}}}\right )\right ],\\
V^{2}_{1{\rm h}{\bf k}}&=&{1\over 2}\left [\alpha_{1{\bf k}}\left (1-{\xi_{\bf k}\over E_{1{\rm h}{\bf k}}}\right )-
\alpha_{3{\bf k}}\left (1-{M_{\bf k}\over E_{1{\rm h}{\bf k}}}\right )\right ],\\
U^{2}_{2{\rm h}{\bf k}}&=&-{1\over 2}\left [\alpha_{2{\bf k}}\left (1+{\xi_{\bf k}\over E_{2{\rm h}{\bf k}}}\right )-
\alpha_{3{\bf k}}\left (1+{M_{\bf k}\over E_{2{\rm h}{\bf k}}}\right )\right ],\\
V^{2}_{2{\rm h}{\bf k}}&=&-{1\over 2}\left [\alpha_{2{\bf k}}\left (1-{\xi_{\bf k}\over E_{2{\rm h}{\bf k}}}\right )-
\alpha_{3{\bf k}}\left (1-{M_{\bf k}\over E_{2{\rm h}{\bf k}}}\right )\right ],~~~~~~~
\end{eqnarray}
\end{subequations}
satisfy the sum rule: $U^{2}_{1{\rm h}{\bf k}}+V^{2}_{1{\rm h}{\bf k}}+U^{2}_{2{\rm h}{\bf k}}+V^{2}_{2{\rm h}{\bf k}}=1$, where
$\alpha_{3{\bf k}}=[2\bar{\Delta}_{\rm pg}({\bf k})]^{2}/(E^{2}_{1{\rm h}{\bf k}}-E^{2}_{2{\rm h}{\bf k}})$, and
the corresponding effective normal-state pseudogap $\bar{\Delta}_{\rm pg}({\bf k})$ and energy spectrum $M_{\bf k}$ in Eq.
(\ref{self-energy-2}) can be obtained explicitly in terms of the self-energy $\Sigma^{({\rm h})}_{1}(k)$ in Eq.
(\ref{self-energy-1}) as,
\begin{subequations}\label{pseudogap}
\begin{eqnarray}
\bar{\Delta}_{\rm pg}({\bf k})&=&{L_{2}({\bf k})\over 2\sqrt{L_{1}({\bf k})}},\\
M_{\bf k}&=&{L_{2}({\bf k})\over L_{1}({\bf k})},
\end{eqnarray}
\end{subequations}
with $L_{1}({\bf k})$ and $L_{2}({\bf k})$ are given by,
\begin{subequations}
\begin{eqnarray}
L_{1}({\bf k})&=&{1\over N^{2}}\sum_{{\bf p}{\bf p}'\mu\nu\tau}\Lambda^{2}_{{\bf p}+{\bf p}'+{\bf k}}{B_{{\bf p}'}
B_{{\bf p}'+{\bf p}}\over 8\omega_{\mu{\bf p}'}\omega_{\nu{\bf p}'+{\bf p}}}\nonumber\\
&\times&{A_{\tau}({\bf p}+{\bf k})F_{\mu\nu\tau}({\bf p},{\bf p}',{\bf k})\over [\omega_{\nu{\bf p}'+{\bf p}}-
\omega_{\mu{\bf p}'}-E^{(\tau)}_{{\rm h}{\bf p}+{\bf k}}]^{2}},~~~~~\\
L_{2}({\bf k})&=&{1\over N^{2}}\sum_{{\bf p}{\bf p}'\mu\nu\tau}\Lambda^{2}_{{\bf p}+{\bf p}'+{\bf k}}{B_{{\bf p}'}
B_{{\bf p}'+{\bf p}}\over 8\omega_{\mu{\bf p}'}\omega_{\nu{\bf p}'+{\bf p}}}\nonumber\\
&\times& {A_{\tau}({\bf p}+{\bf k})F_{\mu\nu\tau}({\bf p},{\bf p}',{\bf k})\over \omega_{\nu{\bf p}'+{\bf p}}-
\omega_{\mu{\bf p}'}-E^{(\tau)}_{{\rm h}{\bf p}+{\bf k}}}.
\end{eqnarray}
\end{subequations}
In this case, it then is straightforward to obtain the effective normal-state pseudogap parameter from Eq. (\ref{pseudogap}) as
$\bar{\Delta}_{\rm pg}=(1/N)\sum_{\bf k}\bar{\Delta}_{\rm pg}({\bf k})$.

\begin{figure}[h!]
\includegraphics[scale=0.5]{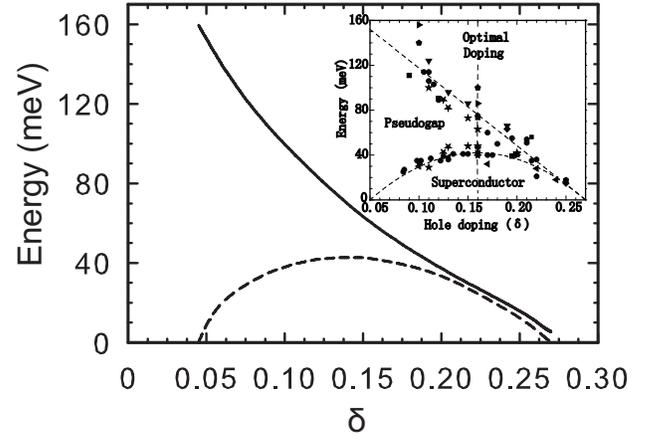}
\caption{The effective normal-state pseudogap parameter (2$\bar{\Delta}_{\rm R}$) (solid line) and effective charge carrier pair
gap parameter ($2\bar{\Delta}_{\rm h}$) (dashed line) as a function of doping for temperature $T=0.002J$ with parameters $t/J=2.5$,
$t'/t=0.3$, and $J=110$meV. Inset: the corresponding experimental data of cuprate superconductors taken from Ref.
\onlinecite{Hufner08}. \label{fig1}}
\end{figure}

Now we are ready to discuss the interplay between the SC gap and normal-state pseudogap. In cuprate superconductors, although the
values of $J$, $t$, and $t'$ are believed to vary somewhat from compound to compound \cite{Damascelli03}, however, as a qualitative
discussion, the commonly used parameters in this paper are chosen as $t/J=2.5$, $t'/t=0.3$, and $J=110$meV. In this case, the
effective normal-state pseudogap parameter (2$\bar{\Delta}_{\rm pg}$) (solid line) and the effective charge carrier pair gap
parameter ($2\bar{\Delta}_{\rm h}$) (dashed line) as a function of doping for temperature $T=0.002J$ are plotted in Fig. \ref{fig1}
in comparison with the corresponding experimental data \cite{Hufner08} of cuprate superconductors (inset). Obviously, the two-gap
feature observed on cuprate superconductors \cite{Hufner08} is qualitatively reproduced. In particular, $\bar{\Delta}_{\rm h}$
increases with increasing the doping concentration in the underdoped regime, and reaches a maximum in the optimal doping, then
decreases in the overdoped regime \cite{feng0306,guo07}. However, in contrast to the case of $\bar{\Delta}_{\rm h}$ in the
underdoped regime, $\bar{\Delta}_{\rm pg}$ smoothly increases with decreasing the doping concentration in the underdoped regime,
this leads to that $\bar{\Delta}_{\rm pg}$ is much larger than $\bar{\Delta}_{\rm h}$ in the underdoped regime. Moreover,
$\bar{\Delta}_{\rm pg}$ seems to merge with $\bar{\Delta}_{\rm h}$ in the overdoped regime, eventually disappearing together with
superconductivity at the doping concentrations larger than $\delta\sim 0.27$.

\begin{figure}[h!]
\includegraphics[scale=0.5]{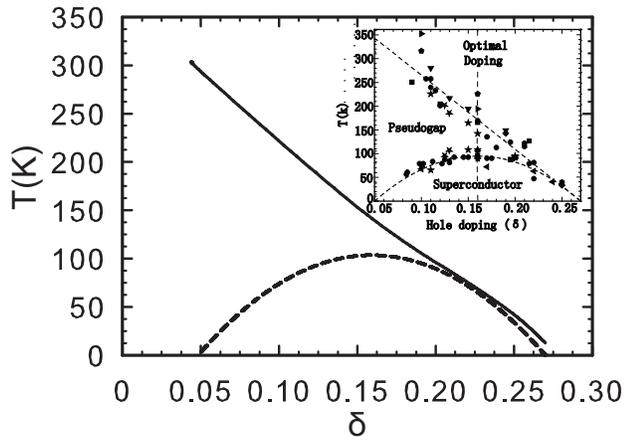}
\caption{The normal-state pseudogap crossover temperature $T^{*}$ (solid line) and superconducting transition temperature
$T_{\rm c}$ (dashed line) as a function of doping for parameters $t/J=2.5$, $t'/t=0.3$, and $J=110$meV. Inset: the corresponding
experimental data of cuprate superconductors taken from Ref. \onlinecite{Hufner08}. \label{fig2}}
\end{figure}

As in the temperature dependence of the SC gap, this normal-state pseudogap is also temperature dependent. In particular, in the
given doping concentration, the normal-state pseudogap vanishes when temperature reaches the normal-state pseudogap crossover
temperature $T^{*}$. To show this doping dependence of $T^{*}$ clearly, we have made a series of calculations for $T^{*}$ at
different doping concentrations, and the results of $T^{*}$ (solid line) and SC transition temperature $T_{\rm c}$ (dashed line)
as a function of doping are plotted in Fig. \ref{fig2} in comparison with the experimental data of cuprate superconductors
\cite{Hufner08}. In corresponding to the results of the doping dependence of $\bar{\Delta}_{\rm h}$ and $\bar{\Delta}_{\rm h}$ as
shown in Fig. \ref{fig1}, $T^{*}$ is much larger than $T_{c}$ in the underdoped regime, then it smoothly decreases with increasing
the doping concentration. Moreover, both $T^{*}$ and $T_{c}$ converge to the end of the SC dome, in qualitative agreement with the
experimental results \cite{Hufner08}.

\begin{figure}[h!]
\includegraphics[scale=0.5]{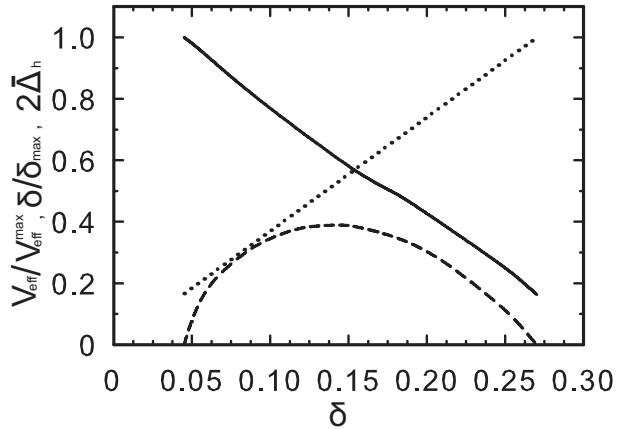}
\caption{The strength of the attractive interaction (solid line), the doping (dotted line), and the effective charge carrier pair
gap parameter ($2\bar{\Delta}_{\rm h}$) (dashed line) as a function of doping for temperature $T=0.002J$ with parameters $t/J=2.5$,
$t'/t=0.3$, and $J=110$meV. \label{fig3}}
\end{figure}

Our present results in Fig. \ref{fig1} and Fig. \ref{fig2} show clearly that there are two coexisting energy gaps in the whole SC
dome: one associated with a direct measure of the binding energy of the two electrons forming a Cooper pair \cite{feng0306}, while
the other with the anomalous normal-state properties \cite{Kastner98,Yang06}. Within the kinetic energy driven SC mechanism
\cite{feng0306}, the essential physics of this two-gap feature in cuprate superconductors can be attributed to the doping and
temperature dependence of the charge carrier interactions in the particle-hole and particle-particle channels directly from the
kinetic energy by exchanging spin excitations. The parent compounds of cuprate superconductors are the Mott insulators
\cite{Damascelli03} as we have mentioned above, when charge carriers are doped into a Mott insulator, there is a gain in the kinetic
energy per charge carrier proportional to $t$ due to hopping, however, at the same time, the magnetic energy is decreased, costing
an energy of approximately $J$ per site, therefore the doped charge carriers into a Mott insulator can be considered as a competition
between the kinetic energy ($\delta t$) and magnetic energy ($J$). This leads to the spin excitation spectral strength decreases with
increasing the doping concentration. In the particle-particle channel, the interaction between the charge carriers mediated by spin
excitations is attractive, then the system of charge carriers forms pairs of bound charge carriers \cite{feng0306}. Since the pairing
force and charge carrier pair gap have been incorporated into the {\it effective} charge carrier pair gap \cite{feng0306} as
mentioned above, the strength $V_{\rm eff}$ of this attractive interaction can be obtained in terms of the {\it effective} charge
carrier pair gap parameter $\bar{\Delta}_{\rm h}$ and charge carrier pair gap parameter $\Delta_{\rm h}$ as
$V_{\rm eff}=\bar{\Delta}_{\rm h}/\Delta_{\rm h}$, where $\Delta_{\rm h}$ is evaluated in terms of the charge carrier off-diagonal
Green's function, and has been given in Refs. \onlinecite{feng0306} and \onlinecite{guo07}. In this case, a decrease of the spin
excitation spectral strength with increasing the doping concentration leads to a decrease of the coupling strength $V_{\rm eff}$ with
increasing the doping concentration. To see this point clearly, we have calculated the doping dependence of $V_{\rm eff}$, and the
results of $V_{\rm eff}/V^{\rm max}_{\rm eff}$ (solid line), $\delta/\delta_{\rm max}$ (dotted line), and $2\bar{\Delta}_{\rm h}$
(dashed line) as a function of doping for $T=0.002J$ are plotted in Fig. \ref{fig3}, where
$V^{\rm max}_{\rm eff}=V_{\rm eff}|_{\delta=0.045}$ is the value of $V_{\rm eff}$ at the starting point of the SC dome, while
$\delta_{\rm max}=0.27$ is the doping concentration at the end point of the SC dome. Our results in Fig. \ref{fig3} show clearly that
the coupling strength $V_{\rm eff}$ smoothly decreases upon increasing the doping concentration from a strong-coupling case in the
underdoped regime to a weak-coupling side in the overdoped regime. Moreover, in the underdoped regime, the coupling strength
$V_{\rm eff}$ is very strong to form the charge carrier pairs for the most charge carriers, and therefore the number of the charge
carrier pairs increases with increasing the doping concentration, which leads to that the charge carrier pair gap parameter and
SC transition temperature increase with increasing the doping concentration. However, in the overdoped regime, the coupling strength
$V_{\rm eff}$ is relatively weak. In this case, not all charge carriers can be bound to form the charge carrier pairs by this weakly
attractive interaction, and therefore the number of the charge carrier pairs decreases with increasing the doping concentration, this
leads to that the charge carrier pair gap parameter and SC transition temperature decrease with increasing the doping concentration.
In particular, the optimal doping is a balance point, where the number of the charge carrier pairs and coupling strength
$V_{\rm eff}$ are optimally matched. This is why the maximal charge carrier pair gap parameter and SC transition temperature occur
around the optimal doping, and then decreases in both the underdoped and overdoped regimes \cite{feng0306}. On the other hand, in the
particle-hole channel, the charge carriers also interact by exchanging spin excitations as in the case in the particle-particle
channel. This interaction in the particle-hole channel reduces the charge carrier quasiparticle bandwidth \cite{guo06}, and therefore
the energy scale of the electron quasiparticle band is controlled by the magnetic interaction $J$. In this case, the effective
normal-state pseudogap parameter $\bar{\Delta}_{\rm pg}$ and normal-state pseudogap crossover temperature $T^{*}$ originated from the
self-energy $\Sigma^{({\rm h})}_{1}(k)$ have the same doping dependent behavior of $V_{\rm eff}$, i.e., they reaches the maximum at
the starting point of the SC dome, and then decreases with increasing the doping concentration, eventually disappearing at the end
point of the SC dome. Furthermore, since the charge carrier interactions in both the particle-hole and particle-particle channels are
mediated by the same spin excitations as shown in Eq. (\ref{self-energy}), therefore all these charge carrier interactions are
controlled by the same magnetic interaction $J$. In this sense, both the normal-state pseudogap and SC gap in the phase diagram of
cuprate superconductors as shown in Fig. \ref{fig1} are dominated by one energy scale. Moreover, our present theory starting from the
$t$-$J$ model also shows that both the normal-state pseudogap and SC gap in cuprate superconductors are the result of the strong
electron correlation.

In conclusion, we have discussed the interplay between the SC gap and normal-state pseudogap in cuprate superconductors based on the
kinetic energy driven SC mechanism. Our results show that the interaction between charge carriers and spins directly from the kinetic
energy by exchanging spin excitations induces the normal-state pseudogap state in the particle-hole channel and SC-state in the
particle-particle channel, therefore there is a coexistence of the SC gap and normal-state pseudogap in the whole SC dome. This
normal-state pseudogap is closely related to the quasiparticle coherent weight, and is a necessary ingredient for superconductivity
in cuprate superconductors. Furthermore, our results also show that both the normal-state pseudogap and SC gap are dominated by one
energy scale, and they are the result of the strong electron correlation.

Although the normal-state pseudogap phenomenon and its relevance to superconductivity can also be discussed starting directly from
some phenomenological theories \cite{Cho06,Millis06,Hufner08b,Norman05,Benfatto00,Yang06,Tesanovic08}, however, we in this paper are
primarily interested in exploring the general notion of the interplay between the SC gap and normal-state pseudogap in the kinetic
energy driven cuprate superconductors. Experimentally, the SC transition is a true transition with all necessary peculiarities in
thermodynamic quantities, whereas the normal-state pseudogap transition is just a crossover, and its effect is reflected in the
anomalous normal-state properties. Recently, the specific-heat measurement \cite{Wen09} on cuprate superconductors shows that the
specific-heat has a humplike anomaly near the SC transition temperature $T_{c}$, and behaves as a long tail which persists far into
the normal-state in the underdoped regime, but in the heavily overdoped regime the anomaly ends sharply just near $T_{c}$. In this
case, we \cite{zhao11} have studied the doping dependence of the thermodynamic properties in cuprate superconductors within the
present framework, and the results show that this humplike anomaly of the specific-heat near SC transition temperature in the
underdoped regime can be attributed to the emergence of the normal-state pseudogap \cite{zhao11}. Furthermore, we \cite{zhao12} have
also discussed the evolution of the Fermi arcs with doping and temperature within the present framework. In particular, we show that
when the temperature $T\sim 0$, the Fermi arc in the underdoped regime is converged around the nodal point, however, with increasing
temperatures, it collapses almost linearly with temperature $T$, in qualitative agreement with the experimental results
\cite{Kanigel06}. These and the related results will be presented elsewhere.

Finally, we have noted that as for the normal-state pseudogap, which grows upon underdoping, it seems natural to seek a connection to
the physics of the AF insulating parent compound \cite{Hufner08}. However, at the half-filling, the t-J model is reduced as the AF
Heisenberg model with an AFLRO. In particular, this AFLRO is kept until the the extreme low doped regime ($\delta <0.045$)
\cite{lee88}. As we have mentioned in Eq. (\ref{cooperpair}), the conduct is disrupted by AFLRO in this extreme low doped regime, and
then our present theory based on the disordered spin liquid state is invalid \cite{feng0306}. In this case, an important issue is how
to extend the present theory in the norma-state for the doped regime without AFLRO to the case in the extreme low doped regime with
AFLRO for a proper description of the connection between the finite doping normal-state pseudogap and the zero-doping quasiparticle
dispersion. These and the related issues are under investigation now.

\acknowledgments

The authors would like to thank Dr. Yu Lan for the help in the numerical calculations. This work was supported by the National
Natural Science Foundation of China under Grant No. 11074023, and the funds from the Ministry of Science and Technology of China
under Grant Nos. 2011CB921700 and 2012CB821403.

\end{document}